\documentclass[pre,preprint,footinbib,a4paper,superscriptaddress]{revtex4}

\usepackage[latin1]{inputenc}
\usepackage{amssymb}
\usepackage{amsmath}
\usepackage{amsbsy}
\usepackage{bm}
\usepackage{graphicx}
\usepackage{verbatim}
\usepackage{color}
\usepackage{hyperref}
\usepackage{cases}

\newcommand*{\Resize}[2]{\resizebox{#1}{!}{$#2$}}%

\begin{document}
\title{Dynamics of elastically strained islands in presence of an anisotropic surface energy}

\author{Guido Schifani}
\email{guido.schifani@inphyni.cnrs.fr}
\author{M\'ed\'eric Argentina}
\author{Thomas Frisch}
\affiliation{Universit\'e  C\^{o}te d'Azur, CNRS,  INPHYNI, Nice, France}

\begin{abstract}
The equilibrium solutions and coarsening dynamics of strained  semi-conductor islands are investigated analytically and numerically. 
We develop an analytical model to study the effect of surface energy anisotropy on the dynamics coarsening of islands. 
We propose a simple model to explain the effect of this anisotropy  on the coarsening time. We find that the anisotropy slows down the coarsening. This effect is rationalised using  a quasi-analytical description of the island profile.
\end{abstract}

\maketitle

\section{Introduction}
The study of elastically strained semi-conductor thin films displays a lot of challenges both from a theoretical and applied point of view \cite{Pimpinelli1998,Politi2000,Ayers2007}. The observations of the self-organized strained islands, which arise on these semi-conductor films, has attracted a lot of interest due to their opto-electronic properties for light emitting diode and quantum dots laser \cite{Shchukin1999,Stangl2004,Brault2016}. In addition, the development of a model that explains the shape and the dynamics of strained islands (quantum dots) remains a fascinating challenge since it involves the dynamical interplay of elastic, capillary, wetting and alloying effects\cite{Spencer1991,Muller2004,Chiu2006,Aqua2013,Aqua2013-PR,Spencer2013,WeiSpencer2016,Rovaris2016,WeiSpencer2017}. 
During the deposition or the annealing of  the semiconductor film in heteroepitaxy on a substrate,    the atomic lattice difference between  the film and the substrate  induces an elastic stress which can lead to a morphological instability \cite{Srolovitz1989,Mo1990,Eaglesham1990,Floro1990,Sutter2000,Tromp2000,Floro2000,Berbezier2009,Misbah2010,Aqua2013-PR}.
This instability, known as the Asaro-Tiller-Grinfeld (ATG)\cite{Asaro1972,Grinfeld1986} instability, leads to the formation of parabolic-shaped islands (prepyramid) \cite{Tersoff2002}, which have been observed in the nucleation less regime \cite{Sutter2000,Tromp2000}. The prepyramids later evolve in pyramids  as more volume is deposited \cite{Tersoff2002}.
Later on, the self-organized strained islands displays a coarsening dynamic for which   it  has been shown theoretically that the surface energy anisotropy   slows down the coarsening \cite{Zhang2000,Aqua2010}.
The cause of the slowing down of the coarsening can be attributed to several effects is still under investigation and  is addressed in this article \cite{Zhang2000,Aqua2010,Korzec2010465,Aqua2013}.

In this present  work, we raise the following question: what is the effect of the amplitude of the surface energy anisotropy on the dynamics of coarsening. We show, using a one-dimensional continuum model and a set of numerical simulations, that coarsening is slowed down as the amplitude of the anisotropy of surface energy increases. Furthermore, we develop a simple model to  quantify the effect of the anisotropy of surface energy on the coarsening time. 
We propose that the main cause of the slowing down of the coarsening is due to the effect of the anisotropy of surface energy. The main effect of the anisotropy of surface energy is to favorize a specific orientation of the surface.  This leads to an influence on the shape of the island and thus this affect the   distribution of the elastic field in the island.  As we shall show, this modification of the shape of the island leads to a change  in the dependency of the chemical potential  with respect to the island  height $h_0$, and as a consequence this slows down the coarsening dynamics.


We first present the one dimensional dynamical model, which takes into account the anisotropy of surface energy and is based on the resolution of the equation of continuum elasticity. Secondly, we describe analytically the equilibrium shape of one dimensional pyramidal island. 
From our model, we estimate the dependency of the chemical potential as a function of the island height. Thirdly, we use the relations obtained in the previous part to propose a simple dynamical model in order to explain how the coarsening time of two anisotropic strained islands increases as a function of the anisotropy strength. 
We conclude our article by illustrating our results with the numerical simulations of the coarsening of an array of islands in the presence of anisotropy.


\section{Continuum model}

Semiconductors film dynamics can be modelled by a mass conservation equation which takes into account the surface diffusion. This surface diffusion current is proportional to gradients of the surface chemical potential $\mu$. In the absence of evaporation the $1D$ equation for the top surface of the film $h(x,t)$ reads:
\begin{equation}
\frac{\partial h}{\partial t} = \mathcal{D}  \sqrt{1+ h_x^2}  \frac{\partial^2 \mu}{\partial s ^2}\ , 
 \label{eqgeneral}
\end{equation}
where $\mathcal{D}$ is the diffusion coefficient, $h_x$ is the slope of the surface height $\partial_x h(x,t)$ and $\partial/\partial s$ the surface gradient \citep{Levine2007,Schifani2016}. 

The chemical potential $\mu$ at the surface is defined by:
\begin{equation}
\mu=\delta \mathcal{F}/\delta h \, .
\label{eq:mu1}
\end{equation}
Here $\mathcal{F}$ is the free energy of the system which encompasses the surface and the elastic contribution $\mathcal{F}=\mathcal{F}_s+\mathcal{F}_{el}$ and $\mu=\mu_s+\mu_{el}$. The surface energy reads
\begin{equation}
\mathcal{F}_s=\int \gamma(h,h_x) \sqrt{1+|h_x|^2}dx .
\label{eq:Fs}
\end{equation}
It includes both wetting effects and surface energy anisotropy. 
 The elastic energy is given by the integration over both the film and the substrate of the elastic energy density, it reads:
\begin{equation}
\mathcal{F}_{el}=\int_{z<h(x)} \mathcal{E}_{el}(x,z)dxdz.
\end{equation}

\begin{figure}[!ht]
\begin{center}
\includegraphics[width=0.9\columnwidth]{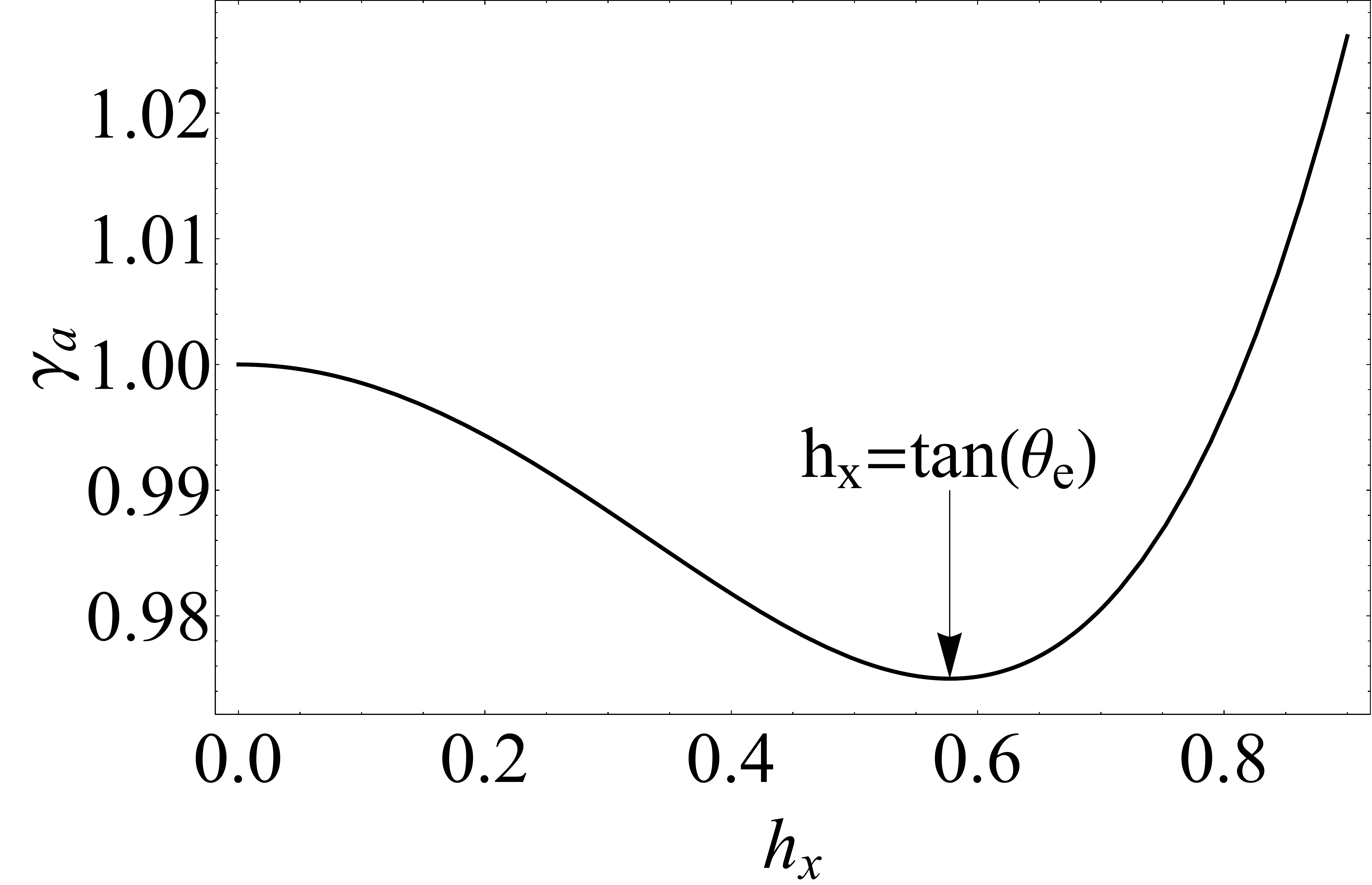}
\caption{\label{fig1} Surface energy  described by the  anisotropy function  given by Eq. (\ref{surf-aniso}) as a function of $h_x$, for $\alpha=0.15$ and $\theta_e=\pi/6$. The minimum is given at $h_x=\tan (\theta _e)$ and represents the characteristic island slope. The vertical axis is in unit of $\gamma_f$.}
\end{center}
\end{figure}

The elastic energy density $\mathcal{E}_{el}$ can be computed using the values of the stress tensor $\sigma_{ij}$ and of the strain tensors $e_{ij}$. It reads,
\begin{equation}
\mathcal{E}_{el} = \frac{1}{2}\sigma_{ij}e_{ij}\,.
\end{equation}
As a first approximation, we examine a decomposition of the surface energy $\gamma(h,h_x)$ where  the  wetting and anisotropic effect are  independent:

\begin{equation}
\gamma(h,h_x)=\gamma_f\left[\gamma_h(h)+\gamma_a(h_x)\right].
\label{eq:gamma-s}
\end{equation}
The wetting effects are linked to the film thickness $h$ through $
\gamma_h (h)=c_w \exp (-h/\delta_w)
$, where $c_w$ and $\delta_w$ are respectively the amplitude and the range of the wetting potential \cite{Muller1996}.
We choose the anisotropy term in the surface energy to have a single minimum at a value $\tan(\theta)=\tan(\theta_e)$, as shown in Fig. \ref{fig1}:
\begin{equation}
\gamma_a	(h_x)=1-\alpha h_x^2\left( 1- \frac{h_x^2}{2\tan^2(\theta_e)}\right)
\, .
\label{surf-aniso}
\end{equation}
Here $\alpha$ is the anisotropy strength and $h_x=\tan(\theta)$  is the surface slope. 
\newpage
Using Eq. (\ref{eq:mu1}) and Eq. (\ref{eq:gamma-s}), the surface  chemical potential $\mu_s$ is found to be: 
\begin{equation}
\begin{aligned}
\mu_{s}=-\left[1-2 \alpha+A(\alpha,\theta_e)h_x^2+B(\alpha,\theta_e)h_x^4 \right] h_{xx}-\frac{c_w}{\delta_w}e^{-h/\delta_w}.
\end{aligned}
\label{eq:mus}
\end{equation}


Here the parameters $A(\alpha,\theta_e)$ and $B(\alpha,\theta_e)$ are defined as
\begin{eqnarray}
A(\alpha,\theta_e)=\frac{3}{2}\left[-1-4\alpha+4\alpha\cot^2(\theta_e) \right]\,,\\
B(\alpha,\theta_e)=\frac{15}{8} \left[1+2\alpha+4 \alpha  \cot ^2(\theta_e) \right]\, .
\end{eqnarray}

The elastic chemical potential can be written as
\begin{equation}
\mu_{el}=-\mathcal{H}(h_x)\, .
\label{eq:muel}
\end{equation}
where $\mathcal{H}(h_x)$ is the Hilbert transform of the spatial derivative of $h(x,t)$, defined as $\mathcal{F}^{-1}(|k|\mathcal{F}(h))$, where $\mathcal{F}$ is the Fourier transform \cite{Aqua2007}. Thus the total chemical potential $\mu$ reads,
\begin{equation}
\mu=\mu_s+\mu_{el}\,.
\end{equation}
The evolution equation for the  surface $h(x,t)$  will merely follow from Eq. (\ref{eqgeneral})
and from the expression  of the surface chemical potential  Eq. (\ref{eq:mus}) and of  the elastic chemical potential Eq. (\ref{eq:muel}).
We first consider the space scale 
\begin{equation}
l_0=\gamma_f / [2(1 + \nu)\mathcal{E}_0]\,,
\label{eq:l0}
\end{equation}
resulting from the balance between the typical surface energy $\gamma_f$ and the elastic energy $\mathcal{E}_0$ density.
Here  $\mathcal{E}_0 = E \, \eta^2/(1-\nu)$ where   $\eta = (a_f - a_s)/a_s$ is the misfit  parameter where  $a_f$ (resp. $a_s$) is the film (resp.
substrate) lattice spacing, $E$ is the Young's modulus of the film and the substrate, and $\nu$ the Poisson's coefficient. 
Secondly we consider the time scale 
\begin{equation}
t_0 =l^4_0/(\mathcal{D} \gamma_f)\,,
\label{eq:t0}
\end{equation}
where $\mathcal{D}$ is the surface diffusion coefficient. In units of $l_0$ and $t_0$, the evolution equation reads
\begin{equation}
\frac{\partial h}{\partial t}=\frac{\partial^2 \mu}{\partial x^2}=\frac{\partial^2 (\mu_{s}+\mu_{el})}{\partial x^2}\,,
\label{eq:equil}
\end{equation}
where $\mu_{s}$ and $\mu_{el}$ are given by Eqs. (\ref{eq:mus}) and (\ref{eq:muel}) respectively. 

The   numerical integration of  Eq. (\ref{eq:equil})   is performed using  a pseudo-spectral method as used  in \cite{Aqua2007,Aqua2010} on a periodic domain of size $L_T$. For example, for a ${\rm Si}_{0.75}{\rm Ge}_{0.25}$ film on ${\rm Si}$, we find $l_0=27\, {\rm nm}$ and $t_0=23\,{\rm s}$ at $700^{\circ}\, {\rm C}$ (see \cite{Chason1999} for an estimate of surface diffusion coefficients). Eq. (\ref{eq:equil}) is  parametrised by the wetting constant $c_w$ and $\delta_w$ and the anisotropy constants $\alpha$ and $\theta_e$. It is a non-linear equation and its evolution is dominated by a coarsening phenomenon in which small islands disappear at the benefit of larger islands. The quantity $S$ defined as the surface of the system 
\begin{equation}
S=\int_{-L_T}^{L_T} h(x)dx\,,
\label{eq:surfacem}
\end{equation}
is conserved during the dynamic as a simple consequence of the form of  Eq. (\ref{eq:equil}).

\section{Analytical model and numerical simulations}

In this section, we first study the equilibrium shape of one island. Using a simple \textit{ansatz}, we analytically determine the characteristics parameters of the island such as its size, height and energy. Our predictions are in good agreement with our numerical computation.
We show that there is a smooth transition from parabolic-like to pyramid-like shapes as S  increases.
Secondly, we derive a dynamical model which shows that the influence of the anisotropy strength $\alpha$ is to increase the coarsening time. Finally, we illustrate our article with the numerical  simulations
of an array of islands displaying coarsening.
\begin{figure}[ht]
\begin{center}
\includegraphics[width=0.9\columnwidth]{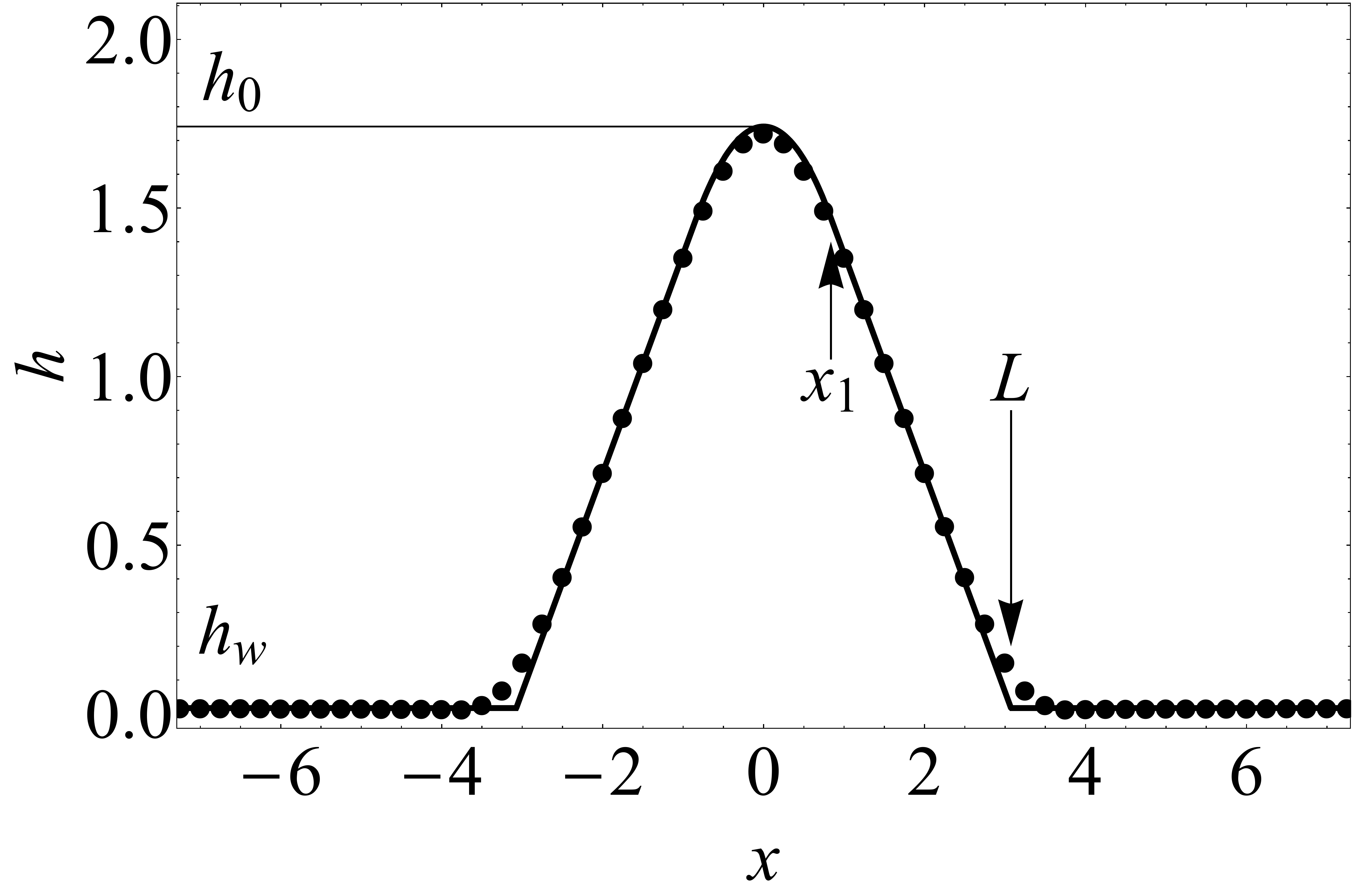}
\caption{\label{fig2} Island like solution given by the numerical simulation of Eq. (\ref{eq:equil}) represented by dots, compared to the  \textit{ansatz} proposed in Eq. (\ref{eq:anzats}) represented by the continuous curve. 
$L$ is the half-width of the island and $x_1$ is the top island parabola width. 
We use as parameters $c_w=0.045$ and $\delta=0.005$. The initial condition is given by a small random perturbation around a constant value of $h=0.4$. The value of the surface is $S=6.1$. The anisotropic parameters are $\alpha=0.1$ and $\theta_e=\pi/6$. The system size is $L_T=16$.}
\end{center}
\end{figure}

\subsection{Equilibrium}
The equilibrium shape of one island is given by the time independent stationary solution of Eq. (\ref{eq:equil}) as shown in Fig. \ref{fig2}.  This solution is characterised by the constant value of the chemical potential $\mu$ along the $x$ axis. The island-like shape can be described using the following \textit{ansatz} for the function $h(x)$:

\begin{eqnarray}
h(x)=
\begin{cases}
 h_w+\tan(\theta_e) (x+L) & -L<x<-x_1  \, , \\
 h_0-\frac{\tan(\theta_e) }{2 x_1} x^2 & -x_1\le x\le x_1  \, , \\
 h_w-\tan(\theta_e) (x-L) & x_1<x<L \, , \\
 h_w & |x|  \geq  L       \, .
\end{cases}.
\label{eq:anzats}
\end{eqnarray}

This pyramidal-like shape describes an island of maximum height $h_0$ which sits on a wetting layer of height $h_w$. 
The island sides are described by straight lines of slope $ \pm \tan(\theta_e)$ for $-L<x<-x_1$ and $x_1<x<L$. 
The island top is described by a parabola for  $-x_1<x< x_1$, which satisfies the continuity of the first derivative  of $h(x)$ at $x=\pm x_1$. 
 At the foot of the island $|x|=L$ the function $h(x)$ is continuous and has a value $h_w$. Here $x_1$ is the half-width of the parabola.
 
 This {\it  ansatz} is characterised by  four  unknown parameters $(h_0, h_w, x_1, L)$ which can be deduced from four relations.
 The first relation is the continuity of $h(x)$ at $x=\pm x_1$, it  imposes:
\begin{equation}
h_0=h_w+\tan (\theta_e)(L-x_1/2)\,.
\label{eq:condh}
\end{equation}

After the substitution of the \textit{ansatz} (\ref{eq:anzats}), we expand  Eq. (\ref{eq:equil}) around $x=0$ in a polynomial series up to second  order in $x$.
At order $0$ in $x$, we obtain the value of the half-width island $L$ 
\begin{equation}
L= \exp{\left(\frac{\pi(1 -2   \alpha) -\pi  \mu  x_1 \cot (\theta_e )+2 x_1( \log (x_1)-1)}{2 x_1}\right)}\,.
\label{eq:condL}
\end{equation}
Here the chemical potential $\mu$  reads:	
\begin{equation}
\mu=-\frac{c_w}{\delta_w}e^{-h_w/\delta_w} \, .
\label{eq:mu-hw}
\end{equation}
This  previous relation  is due to the fact that far from the island the film is flat, so that $h_{x}$ and $h_{xx}$ vanish, and only the wetting potential term remains dominant in Eq. (\ref{eq:mus}) and (\ref{eq:muel}). The chemical potential $\mu$ being fixed, the wetting layer value $h_w$ reads:
\begin{equation}
h_w=-\delta_w \log\left(\mu\frac{\delta_w}{c_w}\right)\,.
\label{eq:mu-hw1}
\end{equation} 

From the expansion at second order in $x$ of Eq.(\ref{eq:equil}), we obtain a transcendental equation for $x_1$, it reads: 
\begin{equation}
\frac{\tan (\theta_e ) \left(\frac{1}{L^2}+\frac{6 \pi  \alpha +x_1}{x_1^3}\right)}{\pi }-\frac{3 (4 \alpha +1) \tan ^3(\theta_e )}{2 x_1^3}=0\,.
\label{eq:condx1}
\end{equation}

Combining Eq. (\ref{eq:condL}) and Eq. (\ref{eq:condx1}), we obtain the following transcendental equation for the parameter $x_1$,
\begin{equation}
\begin{aligned}
&\left(\frac{1}{\pi}\exp \left(-\frac{\pi(1 -2  \alpha) -\pi  \mu  x_1 \cot (\theta_e )+2 x_1(\log (x_1)-1)}
{x_1}\right)+\frac{6 \alpha }{x_1^3}+\frac{1}{\pi  x_1^2}\right)\\
&-\frac{3 (4 \alpha +1) \tan ^2(\theta_e )}{2 x_1^3}=0\,. \end{aligned}
\label{eq:setx1}
\end{equation}


After substitution of Eq. (\ref{eq:mu-hw}) in Eq. (\ref{eq:setx1}), we can solve Eq. (\ref{eq:setx1}) numerically using a simple root finding algorithm to obtain the parameter $x_1$ for different values of $h_w$. 
The island half-width $L$ can then be deduced from Eq. (\ref{eq:condL}). Furthermore the value of the island height $h_0$ can be deduced from Eq. (\ref{eq:condh}). 
For each value of the wetting layer height $h_w$, we can compute the value of the surface $S$ (mass) using Eq. (\ref{eq:surfacem}) and the \textit{ansatz} Eq. (\ref{eq:anzats}).
  Finally, from the knowledge of the values ($h_0$, $L$, $x_1$, $\mu$), we can compute the value of the surface $S$, it reads:
\begin{equation}
\begin{aligned}
S = { } 2 [h_0 x_1+h_w (L_T-x_1)]+\left(3 L^2-6 L x_1+2 x_1^2\right)\frac{\tan (\theta_e )}{3}\,.
\end{aligned}
\label{eq:surface}
\end{equation}

We present in Fig. \ref{fig2} the island shape numerically integrated from Eq. (\ref{eq:equil}) 
and compare it with the \textit{ansatz} (\ref{eq:anzats}). 
 The agreement 
is quite satisfactory and there are no free parameters.

\begin{figure}[!ht]
\begin{center}
\includegraphics[width=0.8\columnwidth]{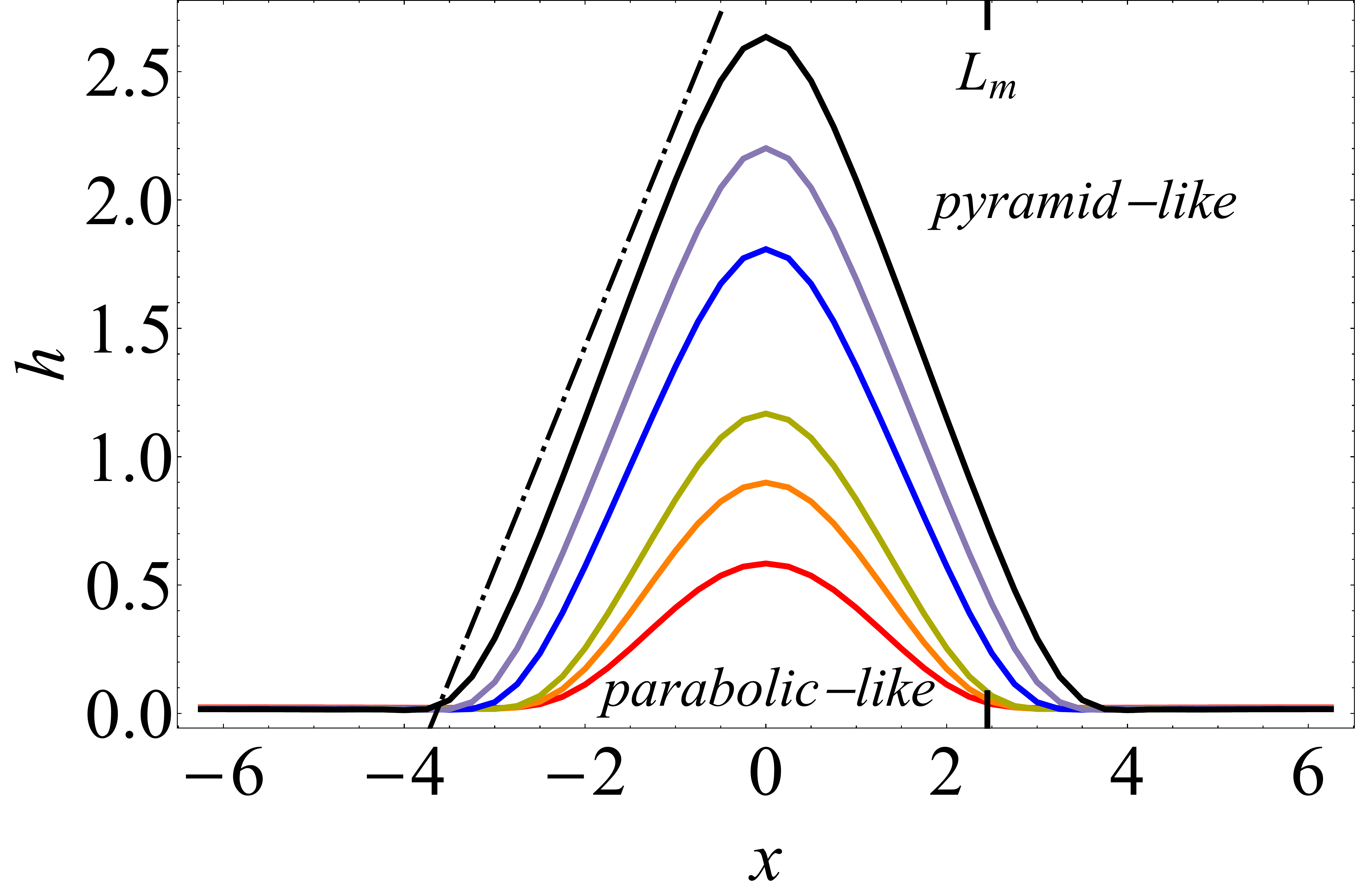}
\caption{\label{fig3} Numerical island profile computed from Eq. (\ref{eq:equil}) for different initial values of the surface $S$. For small surface $S$, we obtain a parabolic-shaped island, with constant island width. For island surfaces larger than $S_m=3.7$, the islands present a pyramidal-like shape. Its characteristic half-width $L_m=2.26$, given by Eq. (\ref{eq:lm}), is represented in the figure. 
From bottom to top: Red-Curve bottom ($S=1.80$), Orange ($S=2.66$), Green ($S=3.48$),  Blue ($S=5.84$) ,  Purple ($S=7.58$), Black ($S=9.74$), color on-line.
The dashed-dotted  curve represent the characteristic slope given by $\tan(\theta_e)$, in this case $\theta_e=\arctan(\sqrt{3/4})$. }
\end{center}
\end{figure}



When the horizontal size of the parabola $2 x_1$  becomes smaller than the island size $L$, the island morphology changes from pyramid-like shape to a parabolic-like shape. Using Eq. (\ref{eq:condL}) and Eq. (\ref{eq:surface}) we obtain  
\begin{equation}
L_m = \frac{3}{5} \pi  \sec ^2(\theta_e ) (1-(8 \alpha +1) \cos (2 \theta_e ))\,,
\label{eq:lm}
\end{equation}
\begin{equation}
S_m = \frac{2}{3} L_m \left[3(h_0+ h_w)+L_m \tan (\theta ) \right]\,.
\label{eq:sm}
\end{equation}

Therefore, the pyramidal shape can only exist for $S>S_m$ and $L>L_m$. Below this value of $S_m$, the islands are parabolic-like shaped and the anisotropy can be neglected.

We plot in Fig. \ref{fig3} various island profiles obtained by numerical simulation for different initial values of the surface $S$. Our numerical simulation shows that for a small island surface $S$, the island shape is parabolic-like and its widths $L$ is rather constant. As the surface of the system increases the islands become pyramid-like and their widths increase smoothly with respect to their height. The transition from parabolic-like shape to pyramid is smooth as the control parameter $S$ is varied.




\begin{figure}[!ht]
\begin{center}
\includegraphics[width=0.9\columnwidth]{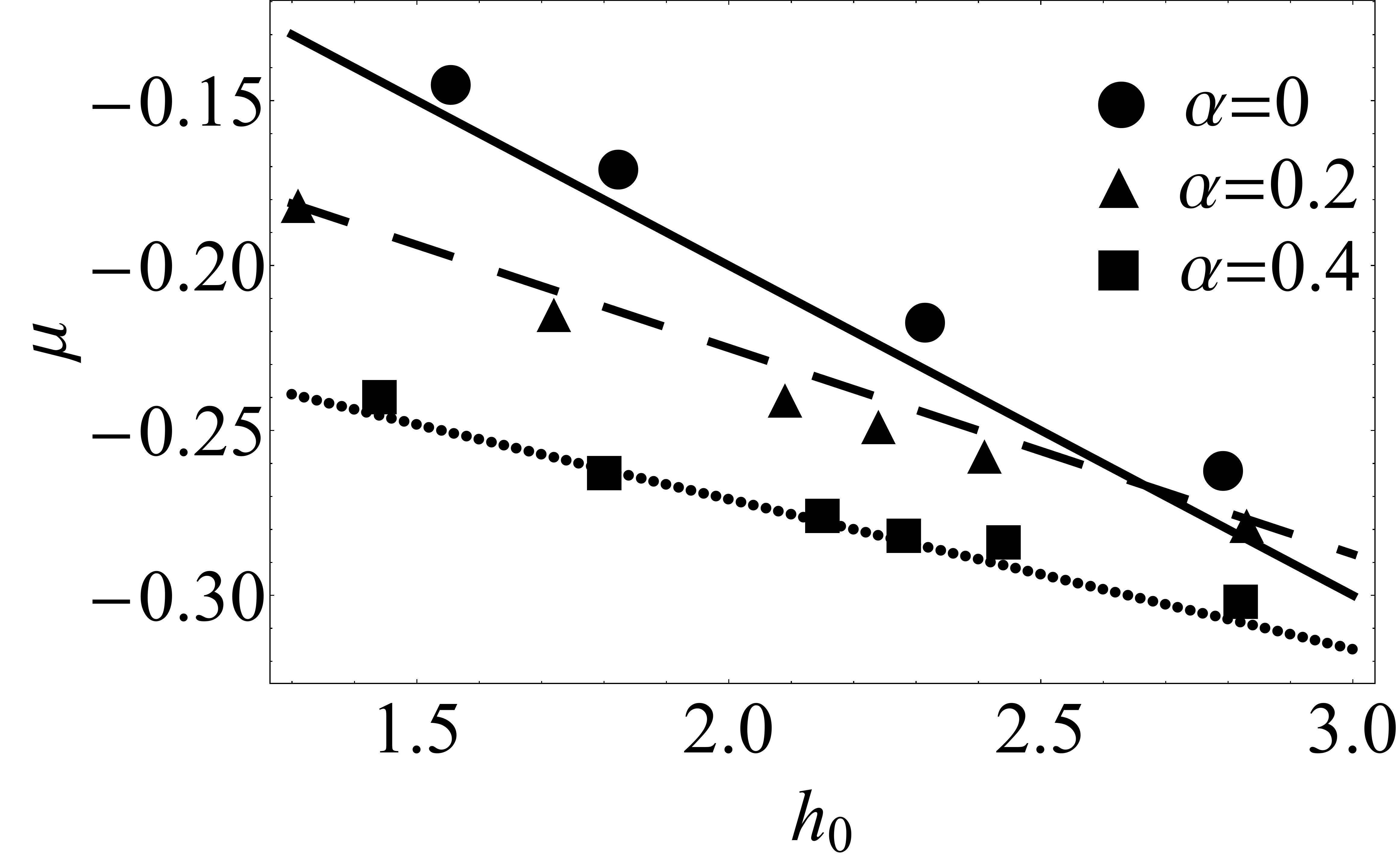}
\caption{\label{fig4} Chemical potential $\mu$ as a function of the island height $h_0$ at equilibrium with $\theta_e=\pi/6$. The $\bullet$, $\blacktriangle$ and $\blacksquare$ 
symbols are computed by numerical integration of Eq. (\ref{eq:equil})
for $\alpha=0$, $\alpha=0.2$ and $\alpha=0.4$ respectively. The continuous, dashed and dotted curves represent the solution given by Eq. (\ref{eq:muhapp}) for $\alpha=0$, $\alpha=0.2$ and $\alpha=0.4$ respectively. The approximation given by Eq. (\ref{eq:muhapp}) fits the results of the numerical solution.}
\end{center}
\end{figure}

Finally, we compute the value of the chemical potential as a function of the island height for various value of the anisotropy strength $\alpha$.
As shown in Fig. \ref{fig4}, the chemical potential decays quasi-linearly as a function of the island height as shown below in Eq. (\ref{eq:muhapp}). 

Using Eq. (\ref{eq:setx1}) the chemical potential $\mu$ can be expressed easily as a function of the parameter $x_1$. In the same way using Eq. (\ref{eq:condh}) the island height can be expressed as a function of $x_1$. Finally using the relation $\frac{\partial \mu}{\partial h}=\frac{\partial\mu}{\partial x_1}\frac{\partial x_1}{\partial h}$ the slope of the chemical potential versus the height of the island for small values of $\alpha$ is found to be:
 \begin{equation}
\partial \mu/\partial h \simeq -\frac{1}{1+3\alpha}\,.
\label{eq:muhapp}
\end{equation}


\newpage

\subsection{Dynamics}
\subsubsection{Numerical simulation of the coarsening of two islands and dynamical model}

\begin{figure}[!ht]
\begin{center}
\includegraphics[width=0.9\columnwidth]{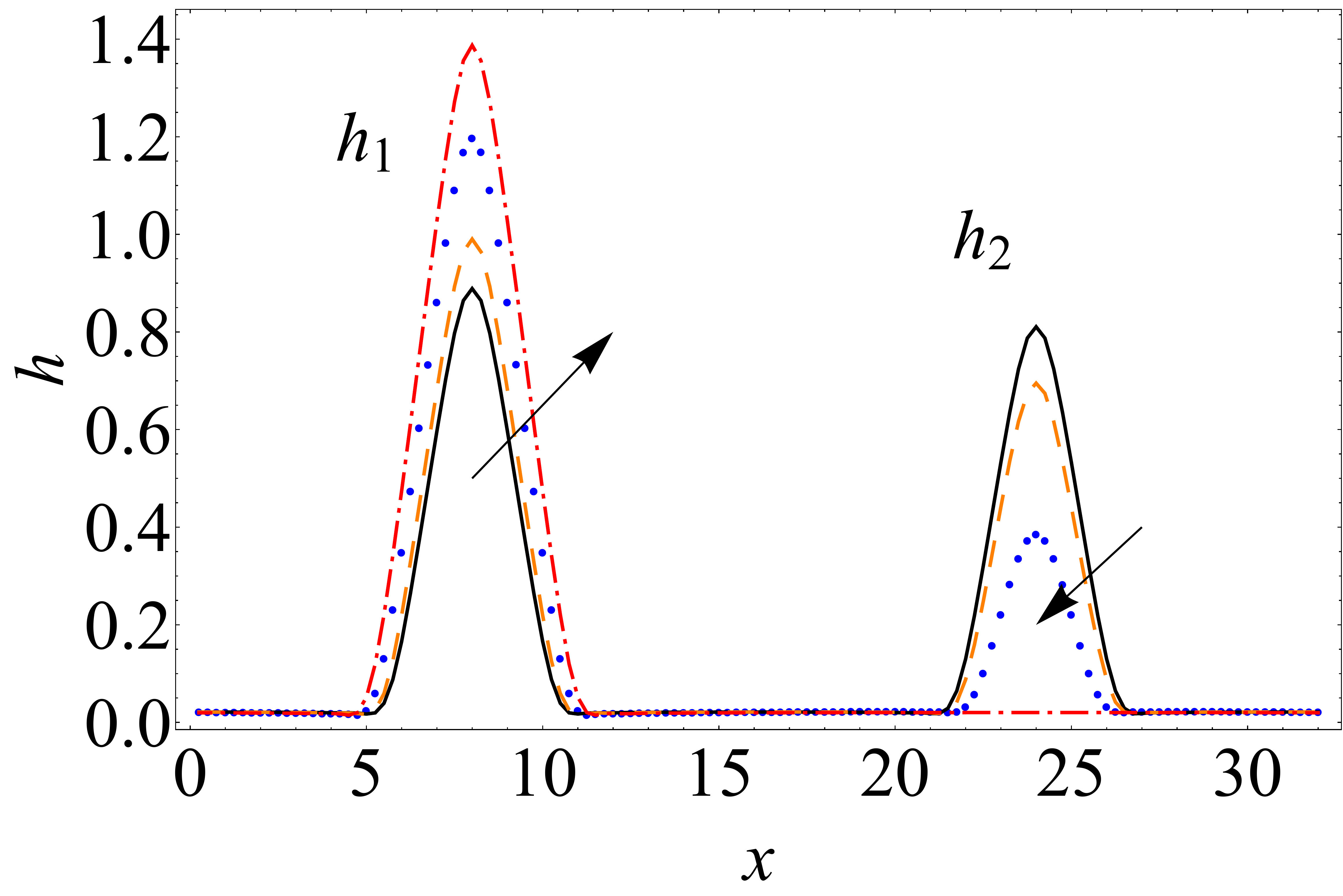}
\caption{\label{fig5} Spatio temporal evolution of two islands, deduced by numerical simulation of Eq. (\ref{eq:equil}). 
The initial condition are two islands of height $h_1(0)=0.89$ and $h_2(0)=0.8$ separated a distance $d=L_T/2$, where $L_T=32$ represents the system size. The anisotropic parameters are $\alpha=0.1$ and $\theta=\pi/6$. The continuous curve (black) represents the island profile at $t=1$, dashed curve (orange) $t=30$, dotted curve (blue) $t=80$ and dotted-dashed curve (red) $t=100$. After a time $t_c=88$, there remains only one island in the system.}
\end{center}
\end{figure}

 In this subsection, we characterise the dynamic of coarsening of two  islands.  In Fig. \ref{fig5}, we show the time evolution of two pyramidal-shaped islands obtained by the numerical simulation of Eq. (\ref{eq:equil}). During the coarsening, the larger island increases at expense of the smaller island until it disappears.
 Ultimately at a time $t_c$ defined as the coarsening time only one island remains in the system. 
  The initial conditions are prepared following \cite{Schifani2016}, by replicating a pyramid with a slight difference in amplitude. 
In order to analyse this phenomenon, we propose a simple model for the coarsening of two islands, inspired by the work presented in \cite{Schifani2016}:
\begin{equation}
 \Resize{5.5cm}{\begin{array}{ll}
h_1\partial_t h_1=\tan(\theta_e)\frac{\mu(h_2)-\mu(h_1)}{d},\\
h_2\partial_t h_2=\tan(\theta_e)\frac{\mu(h_1)-\mu(h_2)}{d}.
\end{array}}
\label{eq:dyn-mod}
\end{equation}
Here $h_1(t)$ is the height of the large island, $h_2(t)$ is the height of the small one and $d$ is the distance separating them. If we consider the island width as $h_i/\tan(\theta_e)$, we recover the model proposed in \cite{Schifani2016}. The advantage of this model is that its resolution requires only the resolution of a differential equation instead of the resolution of a partial differential equation.

\begin{figure}[!ht]
\begin{center}
\includegraphics[width=0.9\columnwidth]{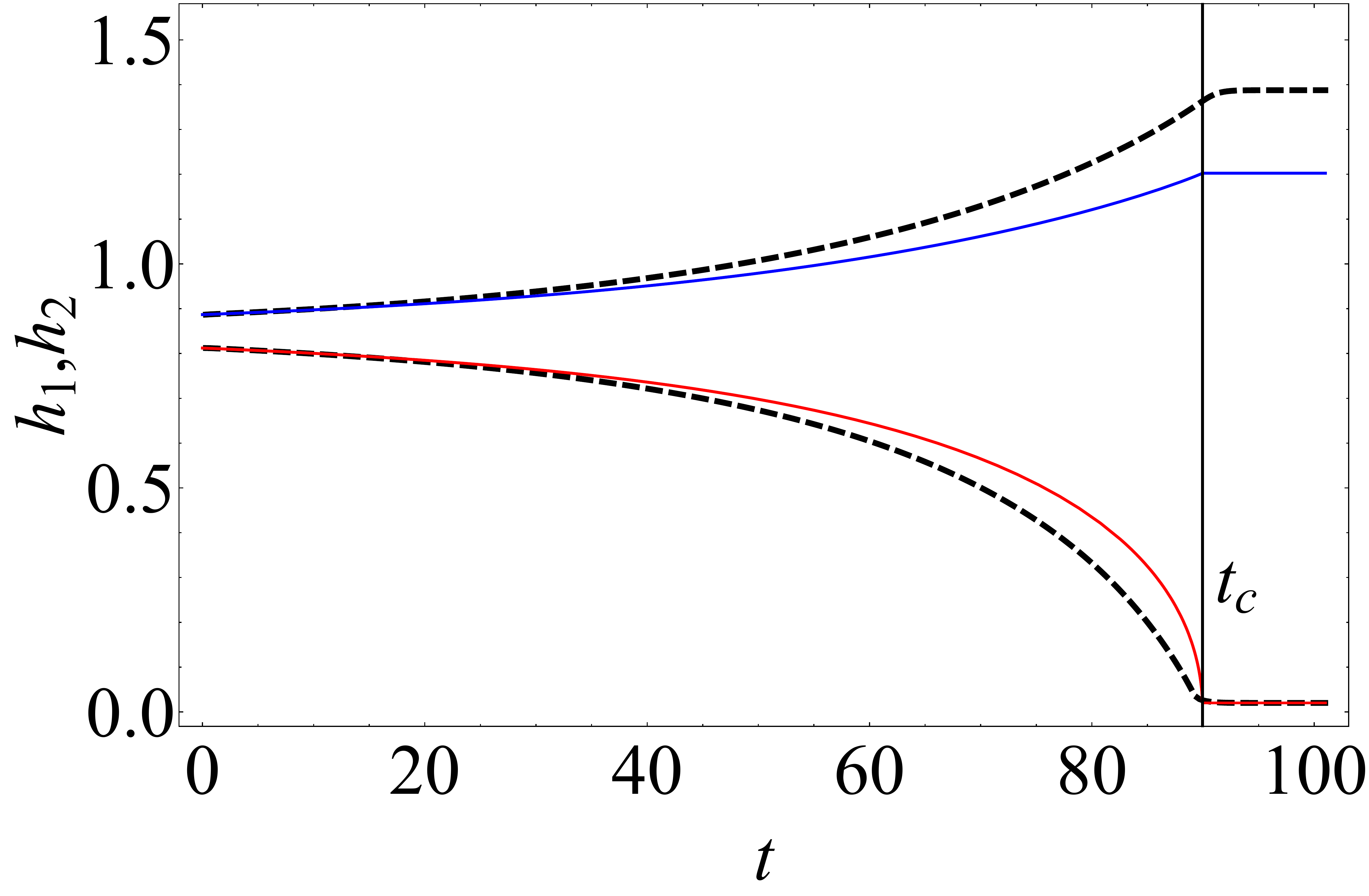}
\caption{\label{fig6} Time evolution of the height $h_1$ and $h_2$, corresponding to the result shown in Fig. \ref{fig5}. The dashed curves represent the islands heights given by the numerical simulation of Eq. (\ref{eq:equil}), and the continuous curves represent the result obtain by the numerical resolution of the dynamical model presented in Eq. (\ref{eq:dyn-mod}). The same results for the continuous curve can be obtained analytical as explained in section $III.B.2$. The coarsening time is $t_c= 88$ for both solutions. The vertical axes is in units of $l_0$ and the time scale is $t_0$.}
\end{center}
\end{figure}

In Fig. \ref{fig6}, we represent the time evolution of each islands heights $h_1$ and $h_2$, corresponding to the result displayed in Fig. \ref{fig5}. We also compare in Fig. \ref{fig6} the results obtained from the resolution of the model Eq. (\ref{eq:dyn-mod}) with results of the numerical simulation of Eq. (\ref{eq:equil}). 
The  model  predictions  is  in good agreement with the numerical simulation of Eq. (\ref{eq:equil}).
The resolution of the model can be done in two ways: a simple numerical integration of Eq. (\ref{eq:dyn-mod}) or an analytical resolution of Eq. (\ref{eq:dyn-mod}) as explain in section $III.B.2$.
The slight discrepancy between the numerical result (numerical simulation of Eq. (\ref{eq:equil})) and theoretical result (resolution of the puntual model Eq. (\ref{eq:dyn-mod})) for the final height $h_1$ 
is mostly due to the fact that our model is based on a simple pyramidal-like shape {\it  ansatz} during all the coarsening dynamic. 
This small discrepancy in the height does not affect the coarsening  time $t_c$. 

\subsubsection{Effect of the anisotropy on the coarsening time}

\begin{figure}[!ht]
\begin{center}
\includegraphics[width=0.9\columnwidth]{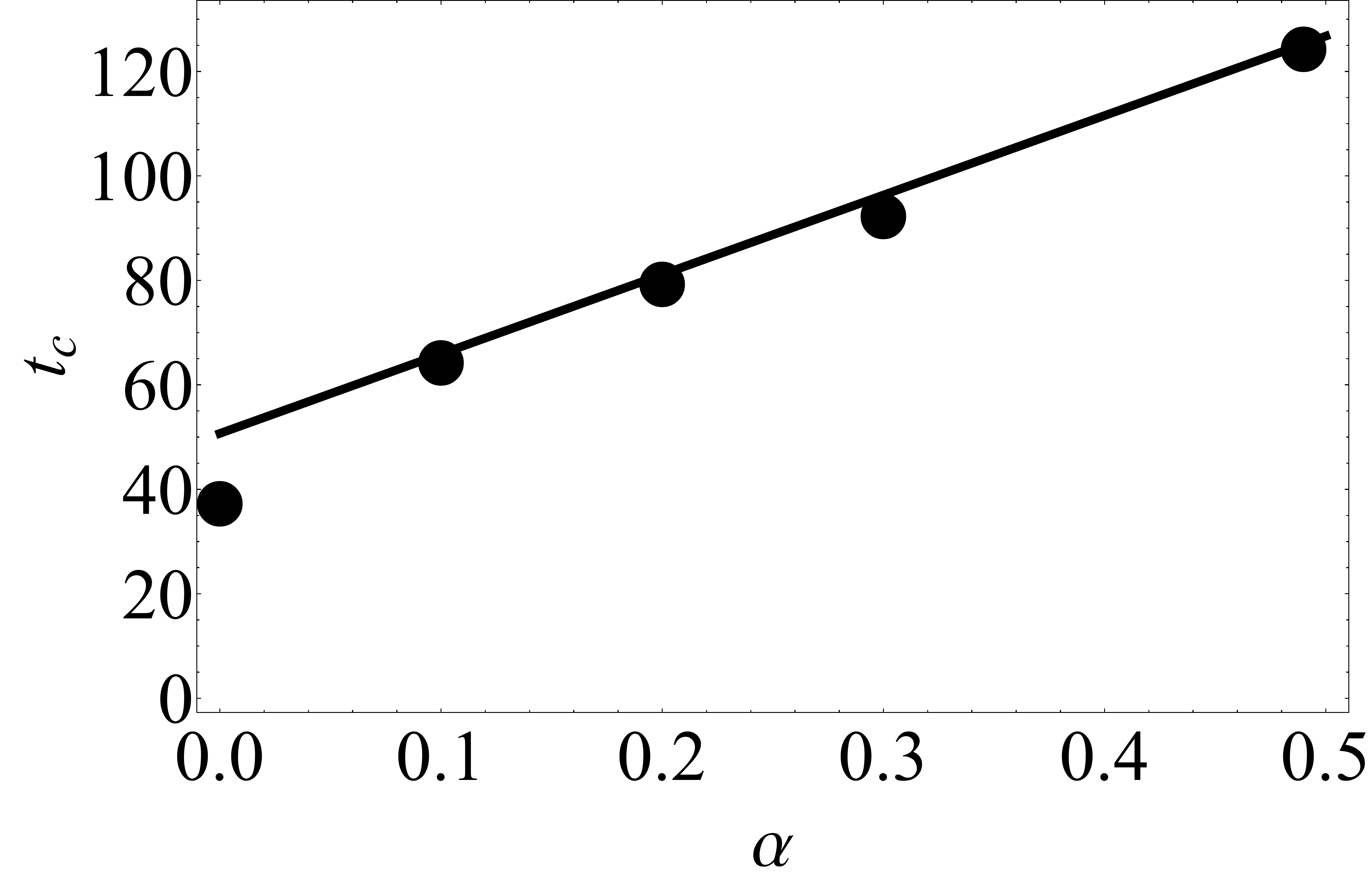}
\caption{\label{fig7} Coarsening time $t_c$ as a function of the anisotropy strength $\alpha$. The system under study is the same as shown in Fig. \ref{fig5}. The dots represents the numerical simulation of Eq. (\ref{eq:equil}) and the curve is the solution for $t_c$ given by Eq. (\ref{eq:tc-analitical}). The coarsening time $t_c$ depends linearly on $\alpha$.}
\end{center}
\end{figure}

In Fig. \ref{fig7}, we present the coarsening time $t_c$ of two strained islands as a function of the anisotropy strength $\alpha$.
We compare the analytical coarsening time $t_c$ obtained by the resolution of the model Eq. (\ref{eq:tc-analitical}) and the results obtained by numerical simulation of Eq. (\ref{eq:equil}). There is a good agreement between both results. As shown in Fig. \ref{fig7} we find that the coarsening time increase linearly as a function of the anisotropy strength $\alpha$. We can explain this effect in the following way.


Using Eq. (\ref{eq:dyn-mod}) it can be easily shown that $\partial_t(h_1^2+h_2^2)=0$. We thus propose the following change of variables in order to solve analytically Eq. (\ref{eq:dyn-mod}):
\begin{equation}
\begin{array}{ll}
h_1(t)=h_0\sin(\phi(t)),\\
h_2(t)=h_0\cos(\phi(t)).
\end{array}
\,
\label{eq:chanvar}
\end{equation}
Here $h_0$ is related to the initial islands heights as $h_0=\sqrt{h_1(0)^2+h_2(0)^2}$. Substituting Eq. (\ref{eq:chanvar}) into Eq. (\ref{eq:dyn-mod}) yields:
\begin{equation}
\partial_t\phi=\frac{\tan(\theta_e)}{h_0d(1+3\alpha)}\left(\frac{1}{\cos(\phi)}-\frac{1}{\sin(\phi)}\right)\, ,
\label{eq:phi}
\end{equation}
submited to the  initial condition $\phi(0)=\phi_0=\arctan\left(\frac{h_1(0)}{h_2(0)}\right)$. 
Eq. (\ref{eq:phi}) can be integrated analytically, its solution is
\begin{eqnarray}
t(\phi)=(1+3 \alpha)d\frac{h_0}{\tan(\theta_e)}  \mathcal{T(\phi)}.
   \label{eq:tvsphi}
\end{eqnarray}
The analitical form of $\mathcal{T}(\phi)$ is given in \footnote{$\mathcal{T}(\phi)=(\sin (\phi )-\sin (\phi_0)+\cos (\phi )-\cos (\phi_0)
-\sqrt{2} \tanh ^{-1}\left[\frac{1}{\sqrt{2}}\left(\tan \left(\frac{\phi
   }{2}\right)+1\right)\right]+\sqrt{2} \tanh ^{-1}\left[\frac{1}{\sqrt{2}}\left(\tan \left(\frac{\phi_0}{2}\right)+1\right)\right])$ where $h_0=\sqrt{h_1(0)^2+h_2(0)^2}$ and $\phi_0=\phi(0)=\arctan\left(\frac{h_1(0)}{h_2(0)}\right)$}.
The coarsening time $t_c$ is defined by the following criteria: when the height of the small island reaches the wetting layer height $h_w$. This implies the following implicit relation for $t_c$:
\begin{equation}
\phi(t_c)=\arcsin\left(\frac{h_w}{h_0}\right).
\end{equation}
This previous relation derives from Eq. (\ref{eq:chanvar}) easily. Using Eq. (\ref{eq:tvsphi}) we obtain the coarsening time $t_c$. It reads:
 \begin{equation}
t_c=(1+3 \alpha)d\frac{h_0}{\tan(\theta_e)}  \mathcal{T}\left[\phi(t_c)\right],
\label{eq:tc-analitical}
\end{equation}
this result agrees with the numerical simulation presented in Fig. \ref{fig7}. In the limit of $(h_w/h_0)\ll1$, the expression $\mathcal{T}(\phi(t_c))$ can be simplified as given in \footnote{$\mathcal{T}(\phi(t_c)) \simeq -\frac{\text{h1}}{\text{h2} \sqrt{\frac{\text{h1}^2}{\text{h2}^2}+1}}-\frac{1}{\sqrt{\frac{\text{h1}^2}{\text{h2}^2}+1}}+\sqrt{2} \tanh ^{-1}\left(\frac{\tan
   \left(\frac{1}{2} \tan ^{-1}\left(\frac{\text{h1}}{\text{h2}}\right)\right)+1}{\sqrt{2}}\right)+1-\sqrt{2} \tanh ^{-1}\left(\frac{1}{\sqrt{2}}\right)$. This result is obtained assuming that the wetting layer height is smaller than the initial islands heights $(h_w/h_0)\ll1$. With this approximation $\phi(t_c)\simeq 0$. }.

\newpage

\subsubsection{Numerical simulation of an array of islands with anisotropy}
\begin{figure}[!ht]
\begin{center}
\includegraphics[width=0.9\columnwidth]{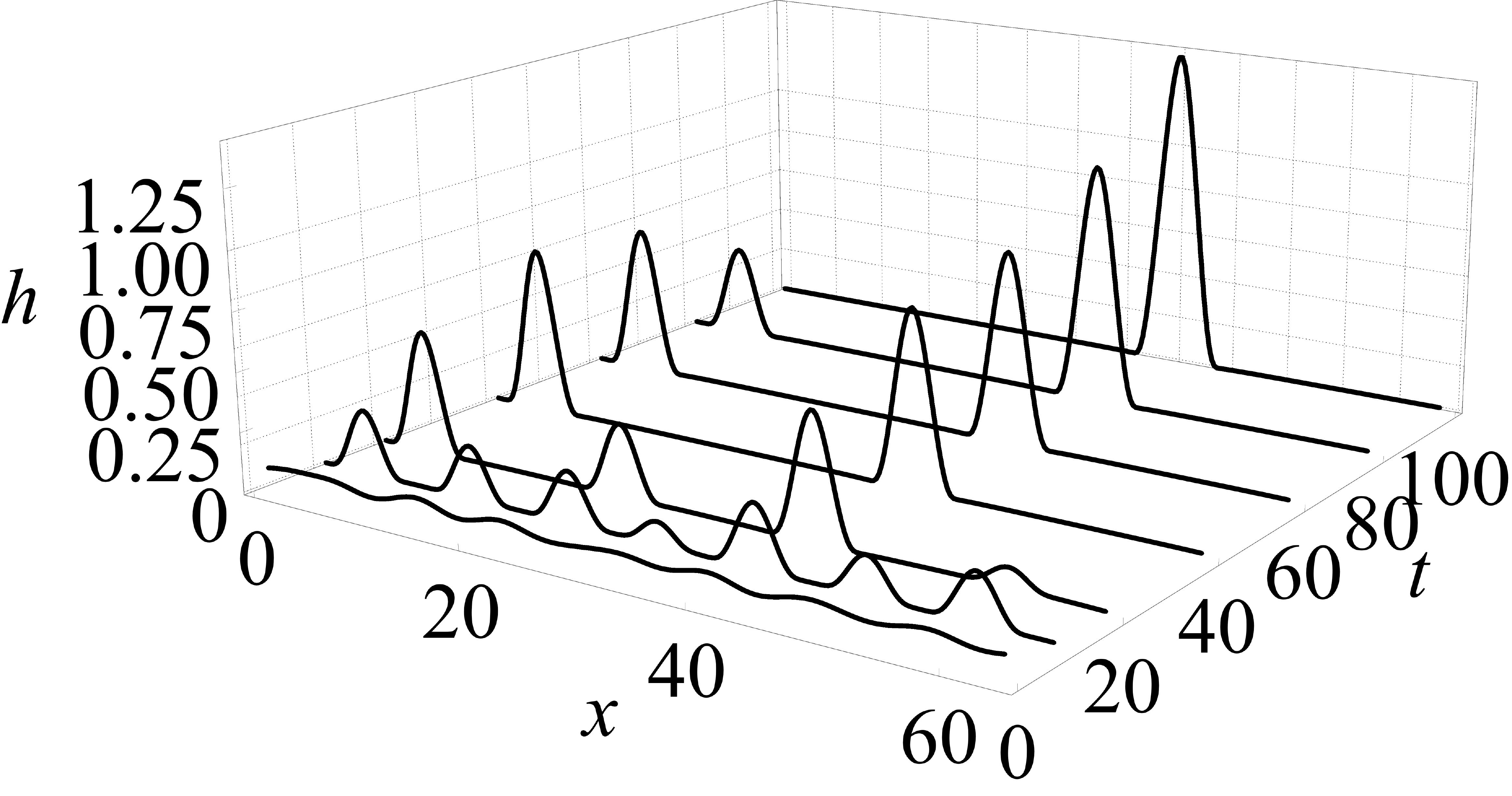}
\caption{\label{fig8} Evolution of an anisotropic strained islands according to Eq. (\ref{eq:equil}), where the initial condition is a flat film of height $0.11$ with a small random perturbation. First the ATG instability evolves, and the coarsening start after the islands have formed. Finally, there is only one pyramidal-shape island with the characteristic slope shown in Fig. \ref{fig1}.}
\end{center}
\end{figure}

For illustration, we present the numerical simulation of the coarsening of an array of islands. The numerical simulation of Eq. (\ref{eq:equil}) reveals mostly two phenomena. A first instability regime which arises for an initial film height higher than the critical layer.
A second regime in which coarsening takes place and is not interrupted. As shown in Fig. \ref{fig8} after the initial instability the smaller islands vanish by surface diffusion through the wetting layer at the benefit of the bigger islands until the system reach the equilibrium. The equilibrium state is characterised by a large island whose characteristic size can be deduced from the parameters of the Eq.(\ref{eq:anzats}). This phenomenon is observed numerically with or without the presence of the surface energy anisotropy. 
\begin{figure}[!ht]
\begin{center}
\includegraphics[width=0.9\columnwidth]{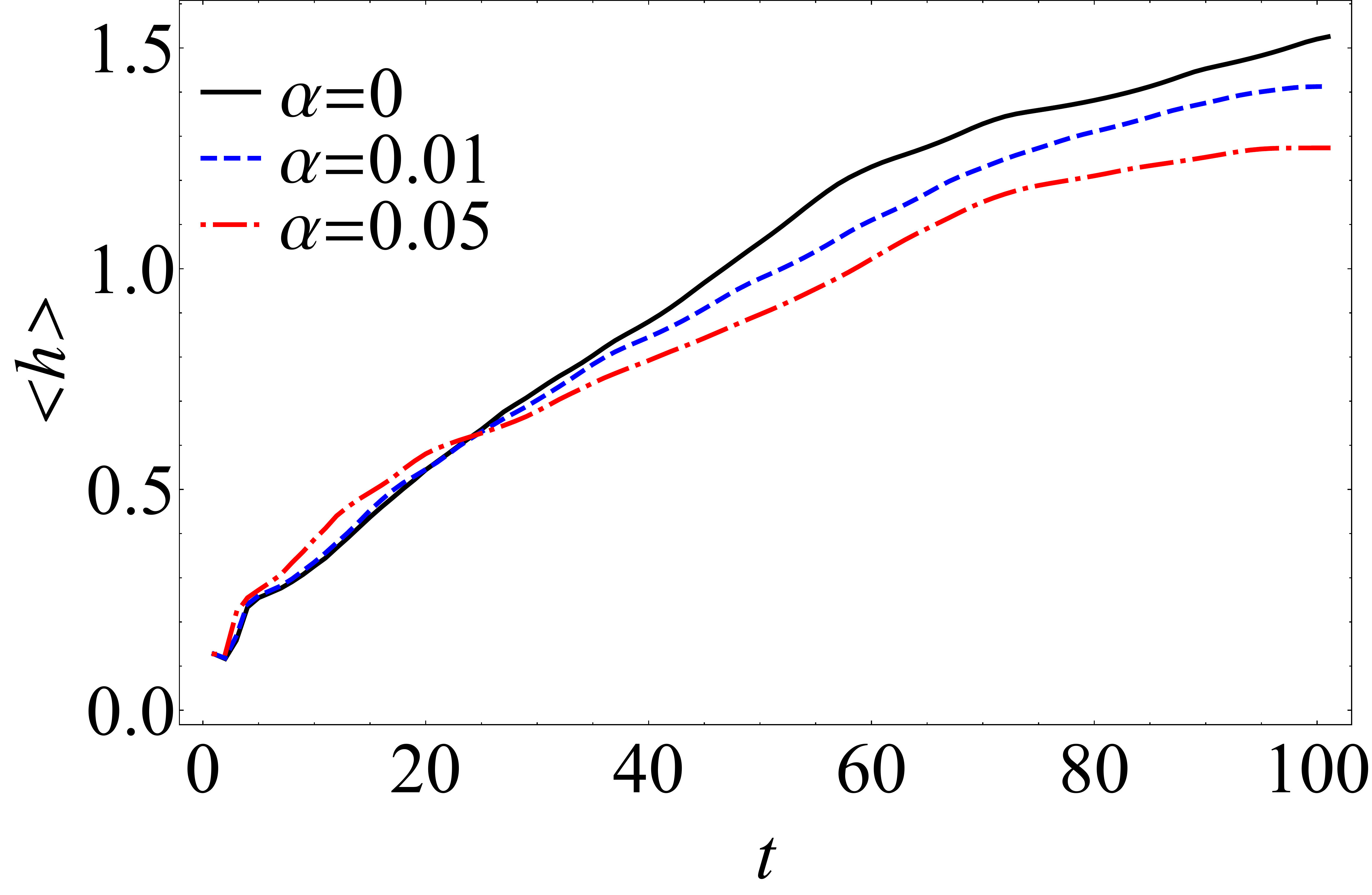}
\caption{\label{fig9}  Ensemble average for the  maximum height $h$ as a function of time  for three different values of $\alpha$. $\alpha=0$ (continuous line),  $\alpha=0.01$ (dashed line) represented in Fig. \ref{fig8}  and  $\alpha=0.05$ (dashed-dotted line) obtain by the numerical simulation of Eq. (\ref{eq:equil}). The initial condition for the systems is a flat film of height $0.1$ with a small random perturbation. The maximum value  of $h$ is calculated using an ensemble average of twenty simulations for each different $\alpha$.}
\end{center}
\end{figure}

We show in Fig. \ref{fig9}  
an ensemble average for the  maximum height  of $h$ as a function of time computed by numerically integrating Eq. (\ref{eq:equil}). The results are presented   for three different values of the anisotropy strength ($\alpha=0$, $\alpha=0.01$ and $\alpha=0.05$). We have performed twenty numerical simulation for each value of $\alpha$.  We observe that the rate coarsening of an anisotropic system ($\alpha=0.01$ and $\alpha=0.05$) is slower than the system without anisotropy ($\alpha=0$). This effect is due to the increase of the coarsening time $t_c$ as described  previously in Fig. \ref{fig7}. Finally, we note that the determination of the coarsening exponent reported in \citep{Aqua2007} is still under investigation in presence or absence of surface anisotropy. As matter of fact the understanding of the dynamics between the islands could serve to elaborate an analytical model to describe the coarsening dynamic of a many islands system.

\newpage
\clearpage

\section{Conclusion}
This article presents a numerical and analytical study of the shape and of the dynamics coarsening of strained anisotropic islands.
We have characterised analytically  strained islands using a simple \textit{ansatz}. We have introduced a dynamical model to investigate the dynamics of coarsening of two islands This models compares favorably with our numerical simulation. 
We have shown that the coarsening dynamics  of strained island in hetero-epitaxy is slowed down by the presence
of the surface energy anisotropy. 
Our results are in good agreement with our numerical simulations. 
For future work the comparison to experiments will be investigated in three dimensions.


\begin{acknowledgements}
We would like to thank Franck Celestini, Jean-No\"el Aqua, Pierre M{\"u}ller and Julien Brault for useful discussions. We thank the ANR NanoGaNUV for financial support.
 \end{acknowledgements}

\end{document}